\magnification=1200 \hsize=6.0 truein
\vsize=8.5 truein
\baselineskip 14pt
\def\ck{1}
\nopagenumbers

\centerline{\bf Erratum}
\vskip 1.0truecm

Fractionally Charged Particles and Supersymmetry Breaking in 4D Strings 
(Phys. Lett. B295 (1992) 219-224) by I. Antoniadis and K. Benakli

In the abstract, ``The communication of the breaking to the observable sector
 is not mediated by
 ordinary gauge interactions." must be replaced by: 
``The communication of the breaking to the observable sector is now mediated by
 ordinary gauge interactions.''

On page 221, in  eq. (4) only the first equation should be kept:

$$
m_{\tilde g_3}= {\alpha_s \over 8 \pi}
4~N~\delta M ~(2 n_Q + n_D) 
$$

In the paragraph below eq. (5), the sentence 
``The remaining one-loop graphs involve the exchange of
gauginos, with masses given in (4), and matter fermions." must be replaced by:
``The leading contribution are two-loop diagrams involving one loop of Standard Model particles and one loop of FEC
particles [\ck]."

Equations (7) must be replaced by:
$$
\eqalign{ 
m_1^2 &=({\alpha_{em} \over  \pi})^2 {1 \over  N^3} {{\delta M^2} \over \cos^4 {\theta_W}} (
n_Q {(2 e_Q-3N)^4 \over 216}\ln ({M_{\rm SU} \over M_Q}) +
n_D {e_D^4 \over 27}\ln ({M_{\rm SU} \over M_D}) \cr & +  
n_L {(2 e_L-N)^4 \over 8}\ln ({M_{\rm SU} \over M_L}) +
 n_E e_E^4 \ln ({M_{\rm SU} \over M_E}))\cr 
m_2^2 &=({\alpha_{em} \over  \pi})^2 N {{\delta M^2} \over \sin^4 {\theta_W}} (
3 n_Q \ln ({M_{\rm SU} \over M_Q}) +  
n_L \ln ({M_{\rm SU} \over M_L})) \cr
m_3^2 &=({\alpha_s \over  \pi})^2 N \delta M^2  (
2 n_Q \ln ({M_{\rm SU} \over M_Q}) +  
n_D \ln ({M_{\rm SU} \over M_D})) \cr}
$$
On pages 223-224, the numerical examples change. As an example consider that:

$$\eqalign{ 
&M_{SU} \sim 3.7 \times 10^{17} {\rm GeV} \quad
\delta M =15 {\rm TeV} \quad {\rm confinement}\quad {\rm group:}\, 
SU(4)\times G,\cr
&n_Q=1, e_Q=1, m_Q=3 \times 10^{14} {\rm GeV},\quad
n_D=2, e_D=1, m_D=2 \times 10^{15}{\rm GeV},\quad \cr
& n_L=0,\quad
n_E=1, e_E=3, m_E= 10^{13} {\rm GeV}. \cr}$$ Then:

\vskip 1.0truecm

\vbox {\tabskip=0pt \offinterlineskip\def\tablerule{\noalign{\hrule}}
\halign to350pt {\strut#& \vrule#\tabskip=1em plus2em& \hfil#& \vrule#&
 \hfil#& \vrule# \tabskip=0pt\cr\tablerule
&&\omit\hidewidth Sparticles \hidewidth&&
 \omit\hidewidth Masses (in GeV) \hidewidth&\cr\tablerule
&&gluinos&&298 &\cr\tablerule
&&charginos&&960 &\cr\tablerule
&& neutralinos&& 320 &\cr\tablerule
&&$\delta m_{\tilde u_L}$&&1350 &\cr\tablerule
&&$\delta m_{\tilde d_L}$&&1350 &\cr\tablerule
&&$\delta m_{\tilde u_R}$&&450 &\cr\tablerule
&&$\delta m_{\tilde d_R}$&&450 &\cr\tablerule
&&$\delta m_{\tilde e_L}$&&1290 &\cr\tablerule
&&$\delta m_{\tilde \nu_L}$&&1290 &\cr\tablerule
&&$\delta m_{\tilde e_R}$&&209 &\cr\tablerule\hfil\cr}}

\item{[{\ck}]} K. Benakli, {\it Quelques aspects de la brisure de la 
supersym\'etrie en th\'eorie des cordes}, Ph.D Thesis, 1994;  M. Dine  
and A. E. Nelson, Phys. Rev. D48 (1993) 1277. \hfill\break

\end